\documentclass[9pt,a4paper]{article}

\usepackage[utf8]{inputenc}

\usepackage[T1]{fontenc}
\usepackage[english]{babel}

\usepackage[scaled]{helvet}

\usepackage{xcolor}

\usepackage{lastpage}
\usepackage{pdfpages}
\usepackage{fancyhdr}

\usepackage{geometry}
\usepackage{setspace}

\usepackage{amsmath}
\usepackage{amsfonts}
\usepackage{amssymb}

\usepackage{graphics}
\usepackage{epsfig}
\usepackage{subfigure}
\usepackage{float}

\usepackage{pdflscape}

\usepackage{array}

\usepackage{overcite}

\usepackage[gen]{eurosym}

\newcommand{\textunderscript}[1]{$_{\text{#1}}$}

\usepackage{caption}
\DeclareCaptionLabelSeparator{pipe}{ \textbf{|} }
\addto\captionsenglish{}

\doublespace

\date{}

\begin{document}

\begin{center}
	\renewcommand{\baselinestretch}{1.1}
	{\large \textbf{The role of lattice mismatch on the emergence of surface states\\
	in 2D hybrid perovskite quantum wells}}

	\vspace{.4cm}

	\normalsize	
	M. Kepenekian,$^{1,*}$ B. Traore,$^1$ J.-C. Blancon,$^2$ L. Pedesseau,$^3$
	H. Tsai,$^{2,4}$ W. Nie,$^2$ C. C. Stoumpos,$^5$
	
	M. G. Kanatzidis,$^5$ J. Even,$^3$ A. D. Mohite,$^2$ S. Tretiak,$^{2,\dagger}$
	and C. Katan$^{1,\ddagger}$\\

	\vspace{.4cm}
	\small\it
	$^1$Univ Rennes, ENSCR, INSA Rennes, CNRS, ISCR
		-- UMR 6226, F-35000 Rennes, France
		
	$^2$Los Alamos National Laboratory, Los Alamos, NM 87545, USA

	$^3$Univ Rennes, INSA Rennes, CNRS, FOTON
		-- UMR 6082, F-35000 Rennes, France

	$^4$Department of Materials Science and Nanoengineering, Rice University,
		Houston, TX 77005, USA
		
	$^5$Department of Chemistry, Northwestern University, Evanston, IL 60208, USA\\
\end{center}

	\vspace{.8cm}

\textbf{%
Surface states are ubiquitous to semiconductors and significantly impact the
physical properties and consequently the performance of optoelectronic devices.
Moreover, surface effects are strongly amplified in lower dimensional systems
such as quantum wells and nanostructures.
Layered halide perovskites (LHPs) are 2D solution-processed natural quantum
wells,~\cite{saparov2016a, pedesseau2016a, li2017a}
where optoelectronic properties can be tuned by varying the perovskite layer
thickness.
They are efficient semiconductors with technologically relevant
stability.~\cite{tsai2016a, yuan2016b, wang2016c, odenthal2017a}
Here, a generic elastic model and electronic structure modelling are applied
to LHPs heterostructures with various layer thickness.
We show that the relaxation of the interface strain is triggered by perovskite
layers above a critical thickness. This leads to the release of the mechanical
energy arising from the lattice mismatch, which nucleates the surface
reorganization and consequently the formation of lower energy edge states.
These states, which are absent in 3D perovskites, dominate the optoelectronic
properties of LHPs and are anticipated to play a crucial role in the design of
LHPs for optoelectronics devices.
}


Surfaces and interfaces are known to play a central part in the performances of
classical semiconductor based devices.~\cite{zwanenburg2013a, delalamo2011a, hwang2012a}
This holds true for the recently emerged halide perovskites.~\cite{haruyama2016a, murali2017a}
The 2D members of the family, layered halide perovskites (LHPs) have superior
photo- and chemo- stability compared to their 3D counterparts. They show strong
promise in high performance optoelectronic devices such as photovoltaics,
field effect transistors, electrically injected light emission and polarized optical
spin injection.~\cite{tsai2016a, mitzi1994a, kagan1999a, yuan2016b, wang2016c, odenthal2017a}
Their properties depend on the number n of MX$_{\text{6}}$ octahedra that
span the perovskite layer (M is a metal, X a halogen).
As in classical semiconductors,~\cite{delalamo2011a} surface and interface
structures can have a strong influence on the properties of LHPs.~\cite{blancon2017a}
While experimental results exists, especially in Ruddlesden-Popper perovskites
(RPPs) of general formula A'$_{\text{2}}$A$_{\text{n-1}}$M$_{\text{n}}$X$_{\text{3n+1}}$
(A and A’ being cations), there is no simple model to predict and control LHP
surface properties.
Here, we show that optical properties of RPPs are decisively impacted by surface
relaxations occurring for structures with n$>$2. We rationalize these features by
considering LHPs as heterostructures built from the n=1 monolayered perovskite
A'$_{\text{2}}$MX$_{\text{4}}$ and the n=$\infty$ 3D AMX$_{\text{3}}$.
This picture leads to understanding of physical phenomena underpinning surface
reconstruction and concomitant modifications of electronic structure, and allows
to formulate the design principles of LHP materials optimized for optoelectronics,
solid-state lighting or photovoltaics. 

\begin{landscape}
\begin{figure*}[p]
	\begin{center}
		\includegraphics[width=0.90\linewidth]{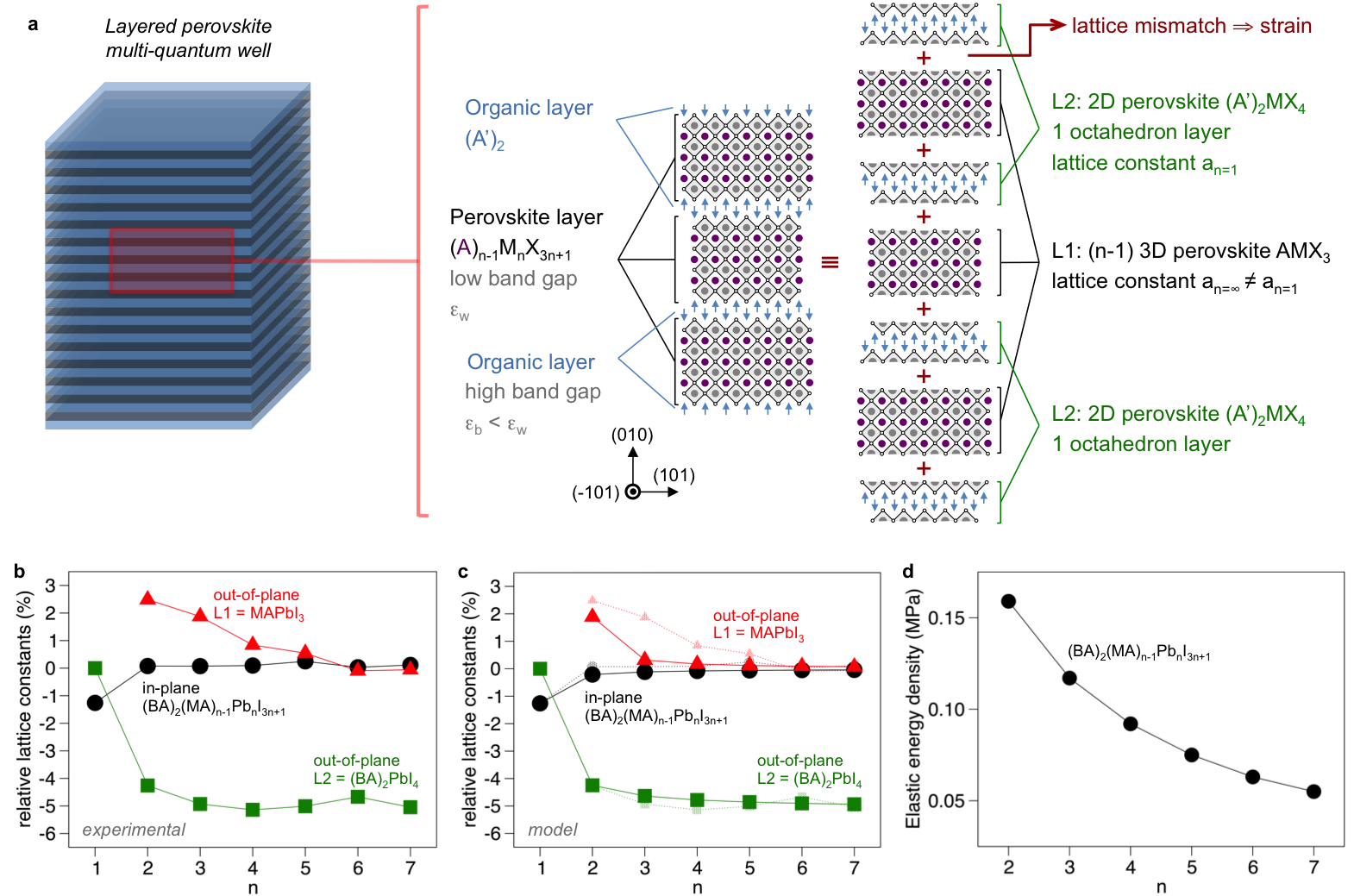}
	\end{center}
	\vspace{-0.2cm}
	\caption{\small
		\textbf{LHPs generalized improper flexoelastic model.}
		\textbf{a}, Schematics of hybrid layered compounds regarded as
			heterostructures L1/L2 with L1 the 3D (n=$\infty$) bulk materials,
			\textit{e.g.} MAPbI$_{\text{3}}$, and L2, a n=1 compound, \textit{e.g.}
			(BA)$_{\text{2}}$PbI$_{\text{4}}$.
		\textbf{b}, In-plane expansion and out-of-plane contractions of
			experimental lattice constants for (BA)$_{\text{2}}$(MA)$_{\text{n-1}}$Pb$_n$I$_{\text{3n+1}}$
			and the L1 and L2 layers.
			The room-temperature structures of MAPbI$_{\text{3}}$ and
			(BA)$_{\text{2}}$PbI$_{\text{4}}$ serve as references for L1 and
			L2 structures, respectively.
		\textbf{c}, Same from the improper flexoelastic model (see \textcolor{orange}{Method}
			for details).
		\textbf{d}, Computed elastic energy density for the 
			(BA)$_{\text{2}}$(MA)$_{\text{n-1}}$Pb$_{\text{n}}$I$_{\text{3n+1}}$ heterostructure.
	}
	\label{fig:elastic}
\end{figure*}
\end{landscape}

We start with evaluating the elastic energy density accumulated in LHPs by
developing an elastic model for the bulk structure
(see \textcolor{orange}{Method} and \textcolor{orange}{Supplementary Information}
for details).
Our model was constructed based on the theory of elasticity in classical
semiconductor heterostructures~\cite{vurgaftman2001a} by identifying the LHP
structure with a multi-quantum well system (\textcolor{orange}{Fig.~\ref{fig:elastic}a})
with alternating stacking of 3D perovskite layers L1 (AMX$_{\text{3}}$, of thickness n-1)
and of 2D perovskite monolayers L2 (single octahedron, n=1).
This combination forms an interface between two structurally-different layers,
equivalent to a so-called L1/L2 heterostructures with a coherent interface
(lattices are continuous across the interface in two directions).~\cite{book_bimberg}
To illustrate this general concept, we consider the family of RPP of general formula
(BA)$_{\text{2}}$(MA)$_{\text{n-1}}$Pb$_{\text{n}}$I$_{\text{3n+1}}$
that can be synthesized in phase-pure form
(only one n-value).~\cite{ruddlesden1957a, ruddlesden1958a, stoumpos2016a}

\textcolor{orange}{Fig.~\ref{fig:elastic}b} represents experimentally observed
variations of the in-plane average lattice parameter as a function of n for the
native (BA)$_{\text{2}}$(MA)$_{\text{n-1}}$Pb$_{\text{n}}$I$_{\text{3n+1}}$
heterostructure, as well as the out-of plane lattice parameters for the
end members of the homologous series L1 ((n-1)MAPbI$_{\text{3}}$) and
L2 ((BA)$_{\text{2}}$PbI$_{\text{4}}$).
As qualitatively predicted based on elasticity, the in-plane lattice expansion from
n=1 to n=$\infty$, gives rise to an out-of-plane lattice contraction in both L1 and
L2 layers.
However, the experimental variation of the in-plane parameter is noticeably steep,
the in-plane parameter of MAPbI$_{\text{3}}$ (n=$\infty$) being almost already
recovered for n=2.
A similar steep variation is observed for the out-plane lattice parameter of the L2
layer.  Interestingly, quantitative agreement between experimental results
and elastic model predictions can only be obtained when a very low effective
stiffness of the L2 layer is considered (\textcolor{orange}{Fig.~\ref{fig:elastic}c}).
The origin of such low stiffness can be traced back to the octahedra tilt angles
extracted from the (BA)$_{\text{2}}$(MA)$_{\text{n-1}}$Pb$_{\text{n}}$I$_{\text{3n+1}}$
RPP experimental structures (\textcolor{orange}{Table \small\bf S1}).
The octahedra tilt angles in L2 structure are indeed more important than in the
L1 layers. In other words, the mechanical energy is more efficiently relaxed in
the RPPs structures by rotation of those octahedra that are directly in contact
with the flexible organic cations, than by Pb-I bond elongation.

Classic theory of elasticity predicts that, for a heterostructure L1/L2 with a
large lattice mismatch between L1 and L2, the structure may undergo a
reorganization for a critical layer thickness, to form nanostructures at the
surface in order to relax the accumulated bulk mechanical energy.~\cite{book_bimberg}
From the above results, the elastic energy density in RPPs with varying perovskite
thickness n was computed (\textcolor{orange}{Fig.~\ref{fig:elastic}d}). We observed
a maximum elastic energy density of $\sim$0.16 MPa for the RPP n=2, and a
monotonic reduction of this energy with increasing n, which ultimately vanishes
for bulk 3D perovskite (n$\rightarrow$$\infty$). Therefore, elastic energy density
arising from the interface is expected to have direct consequences over surface
properties for RPPs with low n-values.

\begin{landscape}
\begin{figure*}[p]
	\begin{center}
		\includegraphics[width=0.85\linewidth]{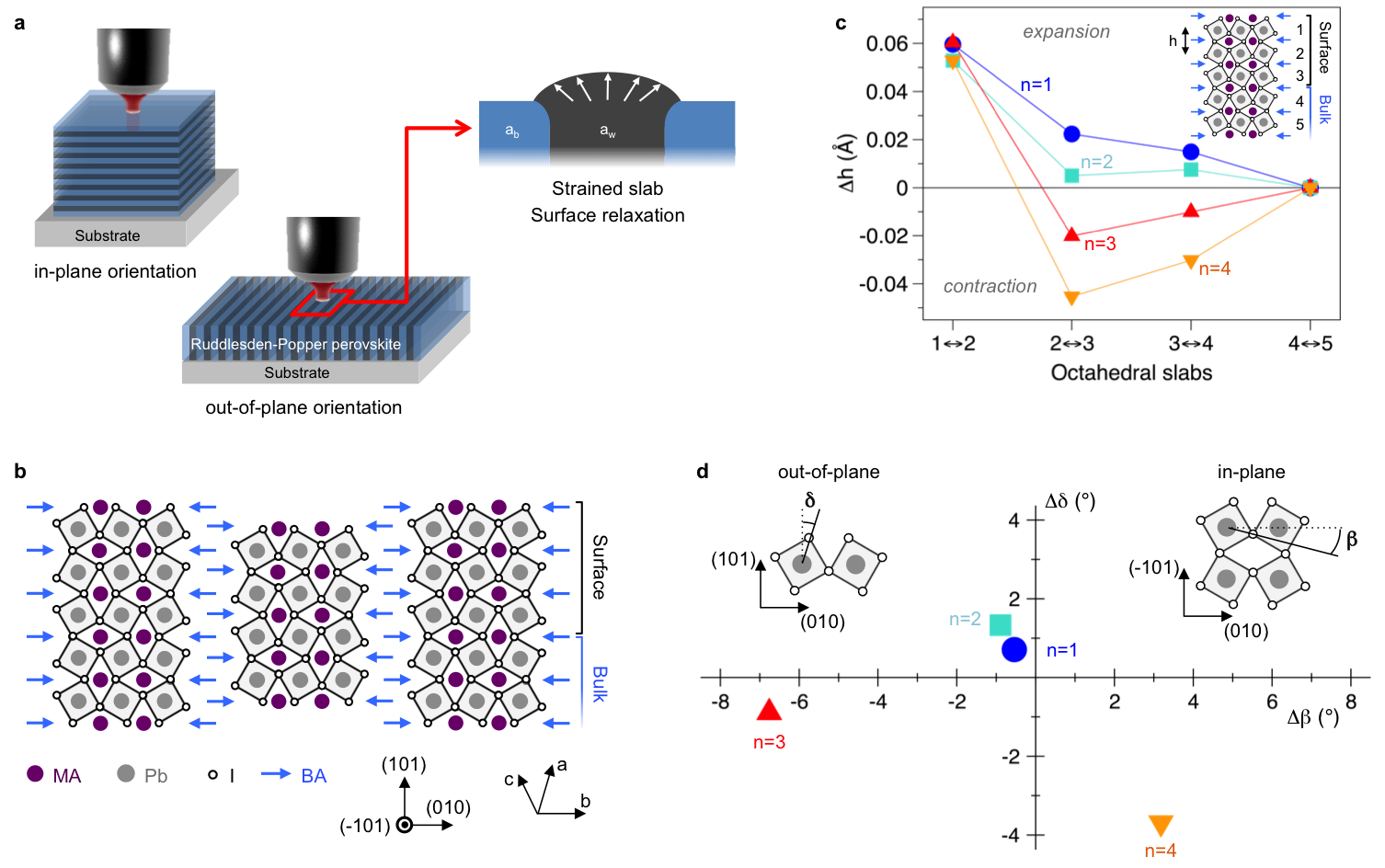}
	\end{center}
	\vspace{-0.5cm}
	\caption{\small
		\textbf{Surface relaxation in LHP multi-quantum wells.}
		\textbf{a}, Schematics of LHP-based devices in in-plane and
			out-of-plane orientation.
		\textbf{b}, Schematics of the (101) surface of the  layered perovskite
			(BA)$_{\text{2}}$(MA)$_{\text{n-1}}$Pb$_{\text{n}}$I$_{\text{3n+1}}$ with n=3.
		\textbf{c}, Variation of the interlayer height difference ($\Delta$h) from
			bulk-like to surface (see inset).
		\textbf{d}, Variation of in-plane ($\beta$) and out-of-plane ($\delta$)
			tiltings of surface octahedra due to the (101) surface relaxation.
	}
	\label{fig:surface}
\end{figure*}
\end{landscape}

We gained further microscopic insight by modelling the structural relaxation at the
relevant surfaces of the (BA)$_{\text{2}}$(MA)$_{\text{n-1}}$Pb$_{\text{n}}$I$_{\text{3n+1}}$
RPP (n=1-4) using density functional theory (DFT), which allows direct simulation
of all structural distortions (see \textcolor{orange}{Method} for computational details).
Applications using RPPs as active materials mainly employ two different
orientations; either in-plane or out-of-plane with respect to the substrate or an
interface layer (\textcolor{orange}{Fig.~\ref{fig:surface}a}). The most relevant
surface of the RPP is then the (101) surface
(\textcolor{orange}{Fig.~\ref{fig:surface}b}),~\cite{tsai2016a, blancon2017a}
which we model here on specifically designed slabs (see \textcolor{orange}{Supplementary Text I}
and \textcolor{orange}{Fig. \small\bf S4}),
labelled as bulk-like and surface, for varying thickness n=1 to 4.
The calculated changes in the surface structure were represented by (i) the
contraction/expansion of the octahedron slabs close to the surface in the
(101) direction (\textcolor{orange}{Fig.~\ref{fig:surface}c}),
and (ii) the in-plane and out-of-plane tilting of the octahedra close to the surface
(\textcolor{orange}{Fig.~\ref{fig:surface}d}).~\cite{pedesseau2016a}
This representation highlights our early conclusion drawn from the elastic model
that rotational degree of freedom of the octahedra play an essential role in relaxing
the internal elastic energy, in contrast with classical semiconductor descriptions
where local strain tensor suffices.~\cite{davies2002a}

\textcolor{orange}{Fig.~\ref{fig:surface}c} shows the variation along the (101)
direction of the distance h between octahedral slabs close to the perovskite
surface. The reference value of h is obtained from the bulk-like region fixed
in our DFT calculations (4$\leftrightarrow$5 distance in \textcolor{orange}{Fig.~\ref{fig:surface}c}).
The evolution of the inter-slab distance yields two opposite behaviours for
n=1,2 and n$>$2.
For n$>$2, the surface slab expansion is accompanied by a contraction of
the sub-surface slabs, which leads to a decoupling of the top surface
octahedron slab from the sub-surface ones.
On the other hand, for n=1,2, expansion of octahedra slabs was observed in
the entire surface region.
A similar distinct behaviour between n=1,2 and n$>$2 was noted by analysing
the surface relaxation in RPPs occurring through in-plane and out-of-plane
tilting of octahedra (\textcolor{orange}{Fig.~\ref{fig:surface}d}).
In fact, surface octahedra in n=1,2 yield almost no rotational degree of freedom,
whereas n$>$2 systems exhibit significantly larger tilting of surface octahedra.
The drastic change of surface behaviour, when increasing the perovskite layer
thickness from n=2 to n=3 attests to a significant change of surface flexibility,
which helps structural relaxation of the internal elastic energy.

\begin{figure*}[p]
	\begin{center}
		\includegraphics[width=1.00\linewidth]{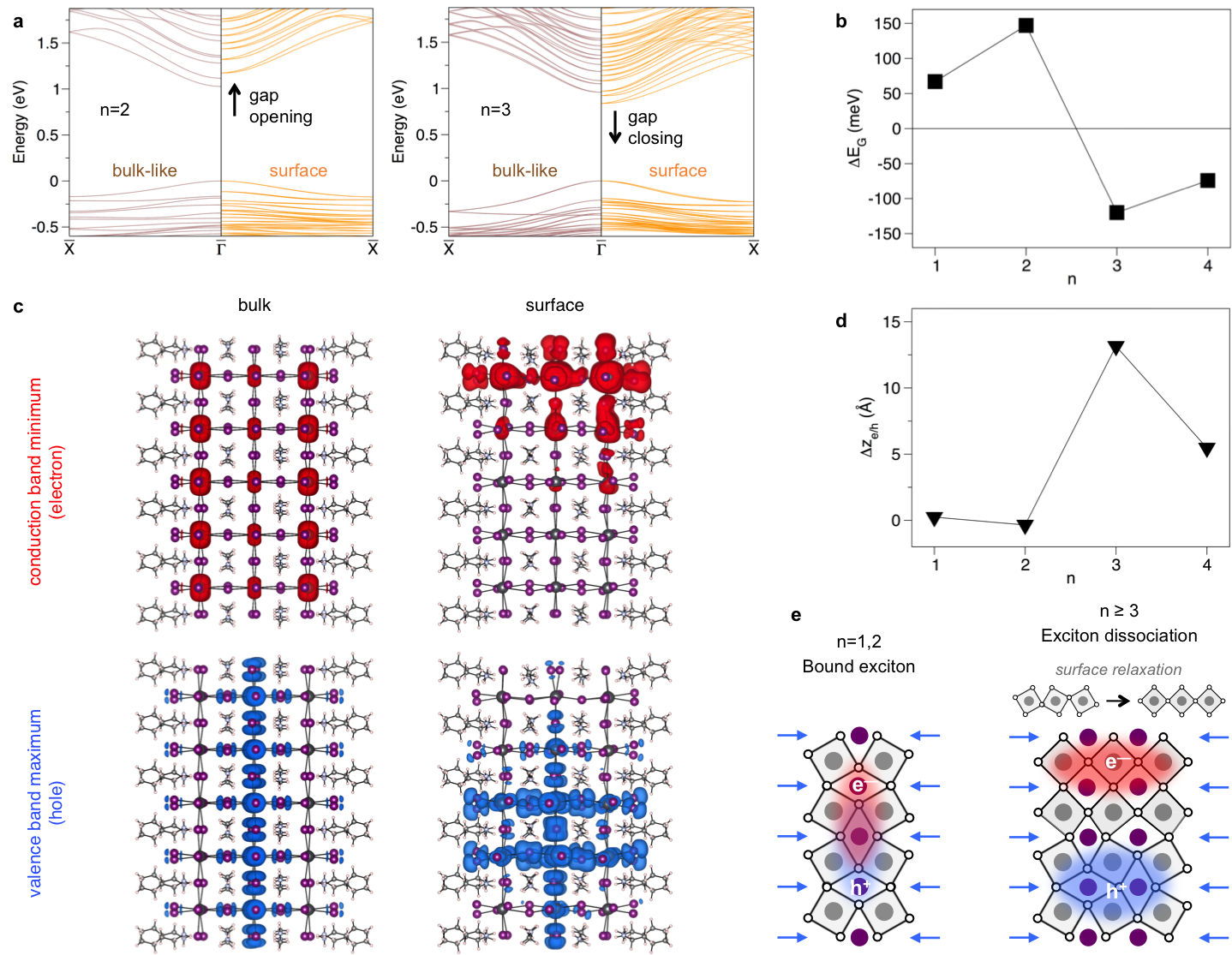}
	\end{center}
	\vspace{-0.2cm}
	\caption{\small
		\textbf{Impact of surface structural relaxation on electronic and optical properties
			in (BA)$_{\text{2}}$(MA)$_{\text{n-1}}$Pb$_{\text{n}}$I$_{\text{3n+1}}$.}
		\textbf{a}, Slab band structures in the bulk-like (left) and relaxed (101)
			surface (right) for n=2 and 3.
		\textbf{b}, DFT variation of E\textunderscript{G} going from bulk-like to relaxed
			(101) surface.
		\textbf{c}, Local densities of states (LDOS) computed at the valence band maximum
			and conduction band minimum for the n=3 RPP in bulk and relaxed surface.
		\textbf{d}, Difference between the barycenter of electron and hole wavefunctions.
		\textbf{e}, Schematics of the surface-induced exciton dissociation
			in RPPs with n$\geqslant$3.
	}
	\label{fig:exciton}
\end{figure*}

We evaluate the impact of these surface relaxation processes on the electronic
and optical properties of RRPs by comparing band structures and wavefunctions
at the surface and in the bulk (\textcolor{orange}{Fig.~\ref{fig:exciton}}).
The electronic band structure still presents a direct bandgap at the surface as
compared to the bulk but with variation of the bandgap energy
(\textcolor{orange}{Fig.~\ref{fig:exciton}a,b} and \textcolor{orange}{Fig. \small\bf S5}).
We observe that the bandgap blueshifts by 70 and 150 meV for n=1 and 2
respectively and redshifts by 120 and 70 meV for n=3 and 4, respectively.
The accuracy of our approach is supported by (i) the excellent agreement
between the calculated exciton properties in the bulk-like region with previously
reported experimental results for the same materials
(see \textcolor{orange}{Supplementary Text II}),~\cite{blancon2017b}
and (ii) the similar pattern in the optical bandgap shift between the RPP layer
surface with respect to the bulk (\textcolor{orange}{Fig. \small\bf S6}).~\cite{blancon2017a}
According to surface relaxation results, lattice expansion at the (101) surface
with relatively small octahedral tilting leads to a bandgap blueshift, whereas
sub-surface lattice compression with significant octahedral distortions results
in a redshift of the bandgap due to appearance of in-gap electronic states.

In order to understand the microscopic impact of the structural changes at
the surface on each type of charges, localized density of states (LDOS) of
the valence band maximum (hole) and conduction band minimum (electron)
were computed (\textcolor{orange}{Fig.~\ref{fig:exciton}c} and \textcolor{orange}{Fig. \small\bf S7a,b}).
For all n-values, surface relaxation leads to hole wavefunctions repelled
away from the surface to the bulk. A similar behaviour is observed for
electrons for n=1,2. In sharp contrast, for n$>$2, the electron gets localized
mainly at the top (101) surface slab. Concomitantly, the preferential direction
of electronic coupling switches from (010) to (101).
From the barycenters of electron (z$_{\text{e}}$) and hole (z$_{\text{h}}$)
LDOS profiles (\textcolor{orange}{Fig. \small\bf S7c,d}), we inspect separation of carriers
($\Delta$z$_{\text{e/h}}$=z$_{\text{e}}-$z$_{\text{h}}$) and demonstrate that
upon appearance of in-gap states, the electron and hole get separated
(\textcolor{orange}{Fig.~\ref{fig:exciton}d}). The effect is maximum for n=3,
$\Delta$z$_{\text{e/h}}$=13.2~\AA{} (5.5~\AA{} for n=4).
Its impact on optical activity is estimated by computing Kane
energies~\cite{kane_book} for bulk-like and relaxed slabs (see \textcolor{orange}{Method}
and \textcolor{orange}{Table \small\bf S3}).
They reflect oscillator strengths of the optical-transitions and show a systematic
reduction by 50\%, 85\%, 30\% and 95\% for the 4 lowest excitations of n=3 RPP
(\textcolor{orange}{Table \small\bf S3}). Such electron-hole separation at the
surface is consistent with the longer photoluminescence lifetime of low-energy
states reported recently.~\cite{blancon2017a}
\textcolor{orange}{Fig.~\ref{fig:exciton}e} summarizes our understanding of
the formation of these low-energy states (LES) in RPPs with n$>$2, which
primarily stems from surface relaxation that strongly localizes the electron
at the surface and facilitates dissociation of the strongly bound bulk exciton.

\begin{figure*}[p]
	\begin{center}
		\includegraphics[width=1.00\linewidth]{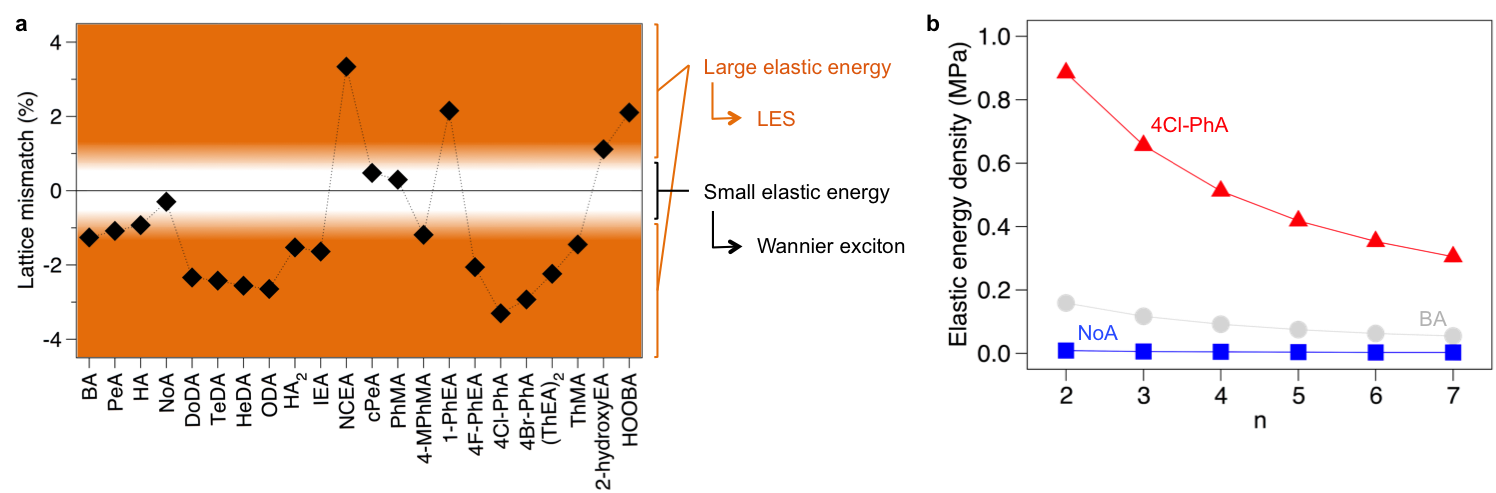}
	\end{center}
	\vspace{-0.2cm}
	\caption{\small
		\textbf{Design of LHPs for photovoltaics and optoelectronics.}
		\textbf{a}, Lattice mismatch between various monolayered
			A'$_{\text{2}}$PbI$_{\text{4}}$ perovskites (n=1) and
			MAPbI$_{\text{3}}$ (\textit{I4cm}; n=$\infty$). All data are taken from X-ray
			structures resolved at room-temperature.
			Names for organic compounds and corresponding references are
			given \textcolor{orange}{Table \small\bf S4}.
		\textbf{b}, Computed elastic energy density for heterostructures built with
			MAPbI$_{\text{3}}$ and (BA)$_{\text{2}}$PbI$_{\text{4}}$ (grey line),
			(C$_{\text{9}}$H$_{\text{19}}$NH$_{\text{3}}$)$_{\text{2}}$PbI$_{\text{4}}$ (NoA, blue line),
			and (4Cl-C$_{\text{6}}$H$_{\text{4}}$NH$_{\text{3}}$)$_{\text{2}}$PbI$_{\text{4}}$ (4Cl-PhA, red line).
	}
	\label{fig:design}
\end{figure*}

LES result from the release of the strain-induced elastic energy at the L1/L2
interface (Fig.~\ref{fig:elastic}).
From our elastic model, the amount of energy accumulated in the materials
is directly dependent on the amplitude of lattice mismatch between layers in
the heterostructure L1/L2 and as a result, tuning the RPP structure and
composition can lead to drastic changes of surface properties.
Using this general approach, the internal elastic energy density accumulated
in the bulk of LHPs can be estimated for any composition and perovskite layer
thickness.
From a practical perspective, understanding the relaxation of the
stored elastic energy at the surface of the LHP materials is of paramount
importance and presents a perfect platform for the systematic and
comprehensive evaluation and screening of LHP compounds with defined
functionalities for novel devices.
This concept is illustrated by changing organic cation A' in RPPs
(\textcolor{orange}{Fig.~\ref{fig:design}a}). For example, replacing BA with
C$_{\text{9}}$H$_{\text{19}}$NH$_{\text{3}}$ (NoA), which has a significantly
smaller lattice mismatch,~\cite{lemmerer2012a} results in the reduction of the
elastic energy density of the RPP composite by more than an order of
magnitude (\textcolor{orange}{Fig.~\ref{fig:design}b}). This would prevent
formation of LES and preserve the bulk Wannier exciton.
By contrast, RPPs based on an organic cation inducing a larger mismatch,
namely (4Cl-C$_6$H$_4$NH$_3$)$_2$PbI$_4$ (4Cl-PhA),~\cite{liu2004a}
undergoes increased strain (\textcolor{orange}{Fig.~\ref{fig:design}a}),
thus larger elastic energy density that should favour significant (101) surface
relaxation suitable for e-h carrier separation.

In summary, we simulated edge (surface) relaxation effects in layered hybrid
perovskite materials and discovered a critical layer thickness above which the
surface reorganization becomes significant. This consequently leads to the
formation of lower energy electronic states rationalizing and confirming
experimental observations.~\cite{blancon2017a}
Our modelling is based on the first generic elastic model for LHPs accounting
for the internal elastic energy accumulated in the material bulk and is further
demonstrated using electronic structure calculations of the surface relaxation
of perovskite layers.
Our observation of electronic bandgap shifts and exciton dissociation at the
surface, depending on the layered perovskite structure distinguishes these
materials from their 3D APbI$_{\text{3}}$ (A=cation; n=$\infty$) counterparts
and pave the way to unique tailored properties and functionalities for
optoelectronic applications.

\clearpage
\newpage
\renewcommand{\baselinestretch}{1.0}
\small


\subsection*{Acknowledgements}
The work in France was supported by Agence Nationale pour la Recherche
(TRANSHYPERO project) and was granted access to the HPC resources of
[TGCC/CINES/IDRIS] under the allocation 2017-A0010907682 made by GENCI.
The work at Los Alamos National Laboratory (LANL) was supported by LANL
LDRD program (J-C.B., W.N., S.T., A.D.M.) and was partially performed at the
Center for Nonlinear Studies. The work was conducted, in part, at the Center
for Integrated Nanotechnologies (CINT), a U.S. Department of Energy, Office
of Science user facility. Work at Northwestern University was supported by
grant SC0012541 from the U.S. Department of Energy, Office of Science.
C.C.S. and M.G.K. acknowledge the support under ONR Grant N00014-17-1-2231.

\subsection*{Author contributions}
M. K., S.T. and C.K. conceived the idea, designed the work, and wrote the
manuscript. J.E developed the semi-empirical BSE approach and the elastic
model. M.K. performed the DFT calculations with support from B.T. and L.P. C.K.
and M.K. analysed the data and provided insight into the mechanisms. M.G.K.
and C.S.S. lend their expertise in chemistry.  A.D.M., J.C., H. T. W. N. supplied
knowledge from an application perspective. All authors contributed to this work,
read the manuscript and agree to its contents, and all data are reported in the
main text and supplemental materials.


\clearpage
\newpage

\clearpage
\newpage



\section*{Supplementary material}

## Methods

**Improper flexoelastic model.** It has been shown earlier that ordered LHPs structures can be treated as composite L1/L2 systems with coherent interfaces, although the definition of the L1 and L2 bulk materials must be chosen with care.[1] Such a framework allows relevant analysis of electronic and dielectric properties of the composite.[1,2] To investigate mechanical properties, a first simple approach may be inspired by the elastic model used to predict the influence of lattice mismatch during epitaxial growth of conventional semiconductor heterostructures having a coherent interface.[3,4] For an epitaxial layer with a low lattice mismatch ($\epsilon_{11}=\epsilon_{22}$) on a rigid (001) zinc-blende substrate, the out of plane strain reads $\epsilon_{33}=(-2C_{12}\epsilon_{11})/C_{11}$. It can be derived using the relation between stress and strain tensors:

$$\begin{pmatrix} \sigma_{11} \\ \sigma_{22} \\ \sigma_{33} \end{pmatrix} = \begin{pmatrix} C_{11} & C_{12} & C_{12} \\ C_{12} & C_{11} & C_{12} \\ C_{12} & C_{12} & C_{11} \end{pmatrix} \begin{pmatrix} \epsilon_{11} \\ \epsilon_{22} \\ \epsilon_{33} \end{pmatrix} \quad (1)$$

and the condition for a free (001) surface ($\sigma_{33}=0$). Alternatively, the same result can be obtained by minimizing the elastic free energy:

$$F = \frac{1}{2} V \left[ C_{11}(\epsilon_{11}^2 + \epsilon_{22}^2 + \epsilon_{33}^2) + 2C_{12}(\epsilon_{11}\epsilon_{22} + \epsilon_{11}\epsilon_{33} + \epsilon_{22}\epsilon_{33}) \right], \quad (2)$$

with respect to $\epsilon_{33}$ ($V$ is the volume of the layer).

This model must be adapted for LHPs. As an example, $(BA)_2MA_{(n-1)}Pb_nI_{(3n+1)}$ can be approximated to an L1/L2 heterostructure with L1=(n-1)MAPbI$_3$ and L2=(BA)$_2$PbI$_4$. Here, the elastic properties of both layers must be taken into account. Considering the transverse elastic approximation for L1=(n-1)MAPbI$_3$:

$$\begin{pmatrix} \sigma_{11,L1} \\ \sigma_{22,L1} \\ \sigma_{33,L1} \end{pmatrix} = \begin{pmatrix} C_{11,L1} & C_{12,L1} & C_{13,L1} \\ C_{12,L1} & C_{11,L1} & C_{13,L1} \\ C_{13,L1} & C_{13,L1} & C_{33,L1} \end{pmatrix} \begin{pmatrix} \epsilon_{11,L1} \\ \epsilon_{22,L1} \\ \epsilon_{33,L1} \end{pmatrix}. \quad (3)$$

Then, the total elastic energy of the L1/L2 composite heterostructure is given by:

$$\begin{aligned} F &= \frac{1}{2} S l_{L1} \left[ C_{11,L1}(\epsilon_{11,L1}^2 + \epsilon_{22,L1}^2 + \epsilon_{33,L1}^2) + 2C_{12,L1}(\epsilon_{11,L1}\epsilon_{22,L1} + \epsilon_{11,L1}\epsilon_{33,L1} + \epsilon_{22,L1}\epsilon_{33,L1}) \right] \\ &+ \frac{1}{2} S l_{L2} \left[ C_{11,L2}(\epsilon_{11,L2}^2 + \epsilon_{22,L2}^2) + C_{33,L2}\epsilon_{33,L2}^2 + 2C_{12,L2}\epsilon_{11,L2}\epsilon_{22,L2} + 2C_{13,L2}(\epsilon_{11,L2}\epsilon_{33,L2} + \epsilon_{22,L2}\epsilon_{33,L2}) \right] \end{aligned} \quad (4)$$

where $S$ is the surface of the coherent interface, $l_{L1}=(n-1)c_{3D}$ and $l_{L2}=c_{2D}$ are the thicknesses of the two layers, $c_{2D}$ and $c_{3D}$ are the stacking parameters of (BA)$_2$PbI$_4$ and MAPbI$_3$, respectively. The two layers share the same

in-plane lattice parameter $a$ leading to the following strain components:

$$\epsilon_{11,L1} = \epsilon_{22,L1} = (a - a_{3D})/a_{3D}$$

and $\epsilon_{11,L2} = \epsilon_{22,L2} = (a - a_{2D})/a_{2D}$.

The conditions for free strain relaxation along the stacking axis lead to:

$$\epsilon_{33,L1} = (-2C_{12,L1}\epsilon_{11,L1})/C_{11,L1}$$

and $\epsilon_{33,L2} = (-2C_{13,L2}\epsilon_{11,L2})/C_{11,L2}$.

Finally, the equilibrium in-plane lattice parameter $a_{eq}$ is derived by minimizing the total energy, yielding:

$$a_{eq} = \frac{\dfrac{K_{L1}}{a_{3D}} + \dfrac{K_{L2}}{a_{2D}}}{\dfrac{K_{L1}}{a_{3D}^2} + \dfrac{K_{L2}}{a_{2D}^2}} \tag{5}$$

with $K_{L1} = (n-1)c_{3D}\left(C_{11,L1} + C_{12,L1} - 2\dfrac{C_{12,L1}^2}{C_{11,L1}}\right)$ and $K_{L2} = c_{2D}\left(C_{11,L2} + C_{12,L2} - 2\dfrac{C_{13,L2}^2}{C_{33,L2}}\right)$.

Noteworthy, this result allows recovering the two limiting cases: for n=1 ($K_{L1}$=0), $a_{eq}$=$a_{2D}$ and for n=∞ ($K_{L1} \gg K_{L2}$), $a_{eq}$=$a_{3D}$. Also, in the special case of a vanishing lattice mismatch ($a_{2D}$=$a_{3D}$), the trivial result $a_{eq}$=$a_{2D}$=$a_{3D}$ is recovered for all n-values.

Unfortunately, measurements of elastic constants in MAPbI$_3$ (L1) remain scarce and reported values come from indirect spectroscopies performed at higher temperatures with large error bars.[5,6] In addition, to the best of our knowledge, no values have been reported for their 2D counterpart (L2). However, we can infer relationships from the nature of both materials as well as from previous experimental results. The tetragonal phase of MAPbI$_3$ can be considered having the same elastic constants on all directions, *i.e* $C_{33,L1} = C_{11,L1}$ and $C_{13,L1} = C_{12,L1}$. (BA)$_2$PbI$_4$ can be regarded as a n=1 slice of MAPbI$_3$. Thus, the in-plane elasticity of L2 is kept identical to the one of L1, *i.e.* $C_{11,L2} = C_{11,L1}$ and $C_{12,L2} = C_{12,L1}$. On the other hand, the out-of-plane elasticity, involving the large organic cations, is expected to be much weaker. Therefore, $C_{33,L2} = C_{33,L1}/2$ and $C_{13,L2} = C_{13,L1}/2$ seems a reasonable choice. As a first guess, we choose $C_{11,L1} = 20$ GPa and $C_{12,L1} = 15$ Gpa.

Meanwhile, we stress that the qualitative trend does not much depend on the precise values of the elastic constants: the model predicts a progressive increase of the in-plane parameter as a function of n, which is underestimated as compared to experimental structural parameters (Fig. S1). The deficiency of such a pure elastic model can be traced back to additional rotational degrees of freedom, namely octahedral tilting, which also afford a path to relax the accumulated mechanical energy. Indeed, it is well known that layered and 3D perovskites undergo both in-plane $\beta_{010}$ and out-of-plane $\delta_{010}$ octahedral tilts (Fig. S2), which have been shown

to affect their electronic properties.[7] It is also known that owing to symmetry properties (halide perovskites are improper ferroelastic materials) the coupling between these rotations and strain components is linear-quadratic.[8] Experimental data (Table S1), show that $\delta_{010}$ of L2 is the most affected for increasing n-values.

Therefore, we modify our model considering that L1 reacts in a purely elastic manner, whereas L2 relaxes the strain through a variation of the $\delta_{010}$ angle, *i.e.* $C_{33,L2} = 0$. Then, the free energy of L2 reads:

$$F_{L2} = Sc_{2D} \left[ C_{11,L2}(\epsilon_{11,L2}^2 + \epsilon_{22,L2}^2) + C_{12,L2}\epsilon_{11,L2}\epsilon_{22,L2} + \frac{1}{2}k_2\delta_{010}^2 + \frac{1}{4}k_4\delta_{010}^4 + \frac{1}{2}\lambda\delta_{010}^2\epsilon_{11,L2} \right]. \quad (6)$$

Again, we use $\epsilon_{11,L2} = \epsilon_{22,L2}$. This leads to an expression similar to (5) for $a_{eq}$:

$$a_{eq} = \frac{\dfrac{K_{L1}}{a_{3D}} - \dfrac{c_{2D}}{a_{2D}}\left(\dfrac{\lambda k_2}{k_4} + \dfrac{\lambda^2}{2k_4}\right)}{\dfrac{K_{L1}}{a_{3D}^2} - c_{2D}\dfrac{\lambda^2}{2k_4 a_{2D}^2}} \quad (7)$$

Once more, physical parameters are needed and available data is currently not sufficient to perform a full translation-rotation mechanical analysis at the same level as the one performed in some 3D perovskite oxides.[9] Neglecting the therm $\lambda k_2/k_4$ allows defining a single $K_{L2}$ parameter from (7), in a similar way than in (5):

$$K_{L2} = c_{2D}\frac{-\lambda^2}{2k_4}.$$

When considering $\lambda^2/(2k_4) = 12$ GPa, the elastic model leads to an excellent agreement with experimental data (Fig. 1b). Again, the qualitative trend does not much depend on the precise values of $\lambda$ and $k$ parameters. For instance, for $\lambda^2/(2k_4)$ ranging from 9 GPa to 16 GPa, the overall conclusions remain valid (Fig. S3).

**Crystal and slab structures.** Calculations of bulk properties are performed in periodic systems using the X-ray resolved geometries for n=1 to 4.[10,11] In these structures, the stacking axis is along the (010) direction. In the devices of interest, the growth direction is not along the stacking axis but along the (101) direction. Thus, before constructing slabs for investigating effect of surface relaxation, we apply an orthorhombic to tetragonal distortion and optimize atomic position of these modified bulk structures. Next, we built 5-octahedron thick slabs oriented in the (101) direction keeping only one out of the two inorganic layers sandwiched between the organic layers to ensure computationally manageable structures. The resulting slabs contain up to 874 atoms. Last, atomic positions of these (101) slabs are relaxed, while keeping the bottom 2 layers frozen in the bulk configuration. Properties of unrelaxed and relaxed slabs are referred to by *bulk-like* and *surface*, respectively. For detailed description of the slab construction see Supplementary Text I.

**Density functional theory calculations.** DFT calculations are conducted with the SIESTA code.[12] The non-local van der Waals density functional of Dion *et al.* corrected by Cooper (C09) is used for geometry optimizations.[13,14] SOC is taken into account through the on-site approximation as proposed by Fernández-Seivane

*et al.*[15] To prevent conflicts between the on-site treatment and the non-locality of C09, single points calculations are conducted with the revPBE functional on which C09 is based. The dipole induced in slabs is treated with the dipole correction as implemented in SIESTA.[16] Core electrons are described with Troullier-Martins pseudopotentials.[17] The valence wavefunction is developed over a double-$\zeta$ polarized basis set of finite-range numerical pseudoatomic orbitals.[18] An energy cutoff of 150 Ry for real-space mesh size is used. Noteworthy, DFT underestimates the bandgap closing upon surface relaxation in RPPs with n>2. Turning to larger supercells, *e.g.* doubled, is likely to allow for additional distortions, which in turn shall further close the bandgap. Unfortunately, such size doubling is computationally too demanding. Besides, for effective masses and complex spinor Bloch functions (Kane model) we used the ABINIT package[19] with GGA-revPBE gradient correction for exchange-correlation,[20] ABINIT projector augmented-wave (PAW) datasets[21] as pseudopotentials for Pb, I, and Cs and an energy cutoff of 19 Ha (517 eV) for the plane-wave basis set. Primitive cells were used for the computation of the electronic band dispersions with Monkhorst-Pack grids for reciprocal space integration of: 4×4×1 for n=1 and 2×2×4 for n=2, 3 and 4. Spin-orbit coupling was included in all calculations. As plane wave calculations are more demanding than the localized basis sets counterpart, organic cations have been substituted by Cs atoms. This substitution is known to leave the band dispersion unaffected close to the Fermi level and, in fact, for n=3 we obtain a comparable bandgap closing (128 meV) and similar spinorial components than with SIESTA and all atoms.

**Localized density of states.** LDOS is a function of energy and position describing the space-resolved density of states (DOS):

$$LDOS(\vec{r}, \epsilon) = \sum_i \langle \Psi_i | \vec{r} \rangle \langle \vec{r} | \Psi_i \rangle \delta(\epsilon - \epsilon_i).$$

Plots are obtained by the integration over a window of energy $\Delta\epsilon$ of 10 meV around the CBM and VBM (Fig. 3c and Fig. S7a,b). Profiles $\rho(z)$ (Fig. S7c), where z is along the (101) direction, are then obtained through:

$$\rho_{\text{CBM}}(z) = \int_{x,y} \int_{\epsilon_{\text{CBM}}}^{\epsilon_{\text{CBM}} + \Delta\epsilon} LDOS(x, y, z, \epsilon) dx dy d\epsilon,$$
$$\text{and } \rho_{\text{VBM}}(z) = \int_{x,y} \int_{\epsilon_{\text{VBM}} - \Delta\epsilon}^{\epsilon_{\text{VBM}}} LDOS(x, y, z, \epsilon) dx dy d\epsilon.$$

From those, we can compute the position barycenters (Fig. S7d) as:

$$z_{\text{CBM}} = \int_z \frac{z \rho_{\text{CBM}}(z)}{\rho_{\text{tot}}} dz, \text{ and } z_{\text{VBM}} = \int_z \frac{z \rho_{\text{VBM}}(z)}{\rho_{\text{tot}}} dz, \text{ with } \rho_{\text{tot}} = \int_z \rho(z) dz.$$

**Kane model.** Optical activity was investigated using Kane's model, used for classical semiconductors, and

4

compute Kane energies according to:[22,23]

$$E_{P_{ij,k}} = \frac{2}{m_e} |\langle \Psi_{VB,i}| - i\hbar \frac{\partial}{\partial x_k} |\Psi_{CB,j}\rangle|^2,$$

where the term in brackets corresponds to the dipolar matrix element between the $i^{\text{th}}$ valence band and the $j^{\text{th}}$ conduction band, and $x_k$ to a crystal axis. This requires knowledge about the complex spinorial components of each band. Unfortunately, such data are not accessible in the SIESTA package[12] when spin-orbit coupling (SOC) is taken into account and we performed single point calculations with the ABINIT software.[19]

# The role of lattice mismatch on the emergence of surface states in 2D hybrid perovskite quantum wells

## Supplementary information

M. Kepenekian,[1,*] B. Traore,[1] J.-C. Blancon,[2] L. Pedesseau,[3] H. Tsai,[2,4] W. Nie,[2] C. C. Stoumpos,[5] M. G. Kanatzidis,[5] J. Even,[3] A. D. Mohite,[2] S. Tretiak,[2,†] and C. Katan[1,‡]

[1] *Univ Rennes, ENSCR, INSA Rennes, CNRS, ISCR – UMR 6226, F-35000 Rennes, France*
[2] *Los Alamos National Laboratory, Los Alamos, NM 87545, USA*
[3] *Univ Rennes, INSA Rennes, CNRS, FOTON – UMR 6082, F-35000 Rennes, France*
[4] *Department of Materials Science and Nanoengineering, Rice University, Houston, TX 77005, USA*
[5] *Department of Chemistry, Northwestern University, Evanston, IL 60208, USA*

## Elastic model for layered hybrid perovskites

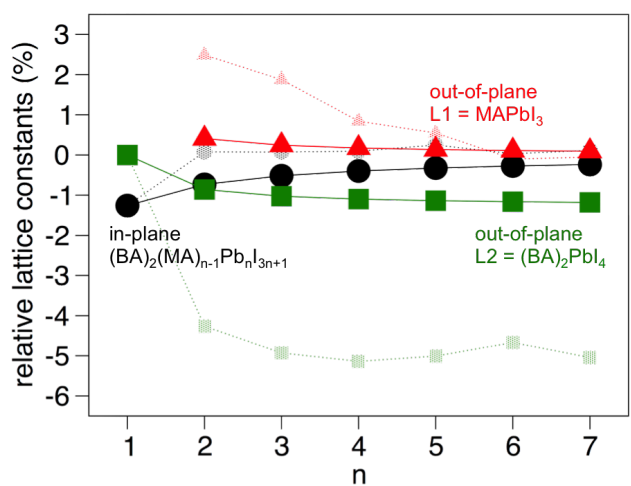

**Figure S1 | Purely elastic model for LHPs.** In-plane expansion and out-of-plane contractions of lattice constants for $(BA)_2(MA)_{n-1}Pb_nI_{3n+1}$ and the L1 and L2 layers from a purely elastic model. The room-temperature structures of $MAPbI_3$ and $(BA)_2PbI_4$ serve as references for L1 and L2 structures, respectively. Shaded symbols correspond to experimental data.

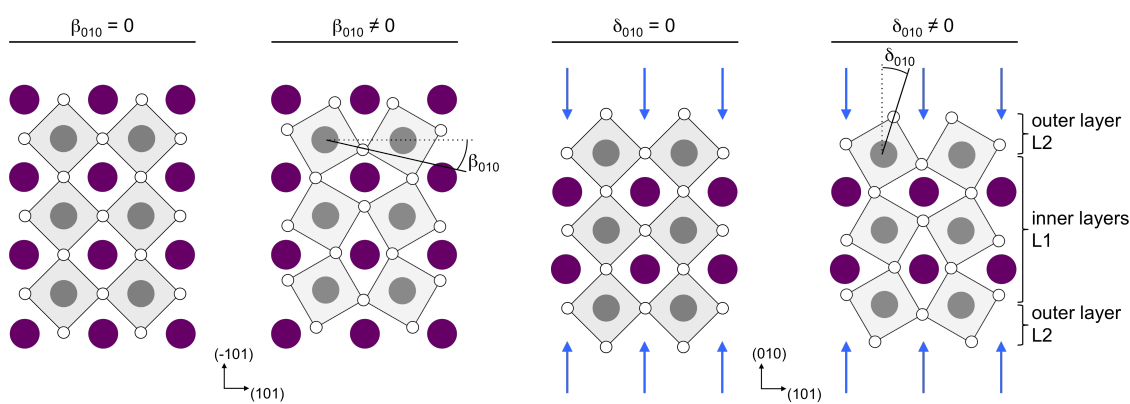

**Figure S2 | Schematic of $\beta_{010}$ and $\delta_{010}$ angles in layered hybrid perovskites for n=3.**

**Table S1 | Layer average $\beta_{010}$ and $\delta_{010}$ angles.** Values are taken from experimental data for the X-ray resolved structures of MAPbI$_3$ (n=∞),[1] (BA)$_2$PbI$_4$ (n=1),[2] and the inner and outer layers of (BA)$_2$(MA)$_{n-1}$Pb$_n$I$_{3n+1}$.[3]

|       | layer | $\beta_{010}$ (°) | $\delta_{010}$ (°) |       | layer | $\beta_{010}$ (°) | $\delta_{010}$ (°) |
|-------|-------|-------------------|--------------------|-------|-------|-------------------|--------------------|
| n=∞   |       | 0.85              | 6.20               | n=5   | outer | 0.00              | 9.95               |
| n=1   |       | 12.27             | 5.78               |       | inner | 0.00              | 6.76               |
| n=2   | outer | 0.00              | 9.93               | n=6   | outer | 0.00              | 10.10              |
|       | inner | 0.00              | 7.18               |       | inner | 0.00              | 5.81               |
| n=3   | outer | 0.00              | 9.67               | n=7   | outer | 0.00              | 10.18              |
|       | inner | 0.00              | 5.23               |       | inner | 0.00              | 7.66               |
| n=4   | outer | 0.00              | 9.95               |       |       |                   |                    |
|       | inner | 0.00              | 6.76               |       |       |                   |                    |

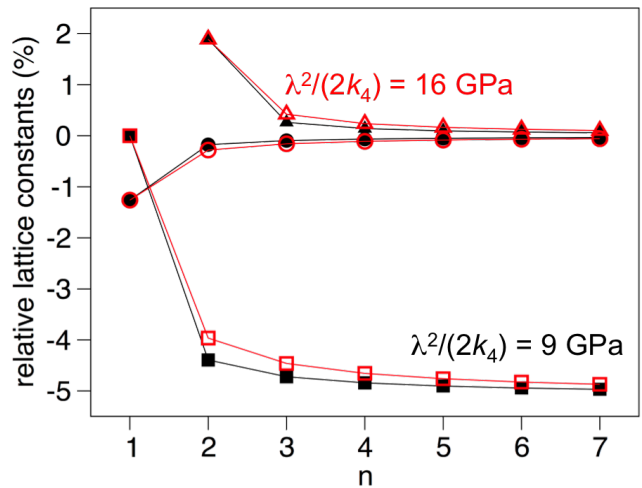

**Figure S3 | Improper flexoelastic model for LHPs.** In-plane expansion and out-of-plane contractions of lattice constants for (BA)$_2$(MA)$_{n-1}$Pb$_n$I$_{3n+1}$ and the L1 and L2 layers from the flexoelastic model using $\lambda^2/(2k_4)$=9 GPa (black symbols) and 16 GPa (red symbols).

## (BA)$_2$(MA)$_{n-1}$Pb$_n$I$_{3n+1}$ RPP (n=1 to 4) from bulk to surface

### Text I | (101) surface construction

Starting from the X-ray structures of (BA)$_2$(MA)$_{n-1}$Pb$_n$I$_{3n+1}$ (BA = butylammonium, MA = methylammonium) with n=1 to 4,[3,4] we apply an orthorhombic to tetragonal distortion, *i.e.* the norms of in-plane cell vectors $a$ and $c$ are averaged to $a' = (a+c)/2$. It leads to a contraction (resp. dilatation) of $a$ (resp. $c$). For n=2 to 4, the distortion is less than 0.5%. It is about 1% for n=1 (Table S2).

DFT relaxation of atomic positions is then performed for these modified bulk structures. Next, starting from the relaxed atomic positions we built slabs oriented in the (101) direction. Because of the orthorhombic to tetragonal transformation, the latter step is straightforward. To reach a computationally manageable structure, we further reduce the number of atoms by keeping one of the two inorganic layers sandwiched between organic layers (Fig. S4). The resulting 5-octahedron thick slabs contain 460, 598, 736 and 874 atoms per cell for n=1, 2, 3 and 4, respectively. Properties of these slabs are denoted as *bulk-like* properties. Worth noticing, all slabs present a global defect of 2 electrons. Along self-consistent calculations, it is compensated by a background charge density.

Then, to describe the surface, atomic positions of these (101) slabs are relaxed, while keeping the bottom 2 layers frozen in the bulk configuration. Properties of these slabs are denoted as *surface* properties. Noteworthy, considering also 4-octahedron thick slabs, we verified that the limited thickness does not alter our conclusions.

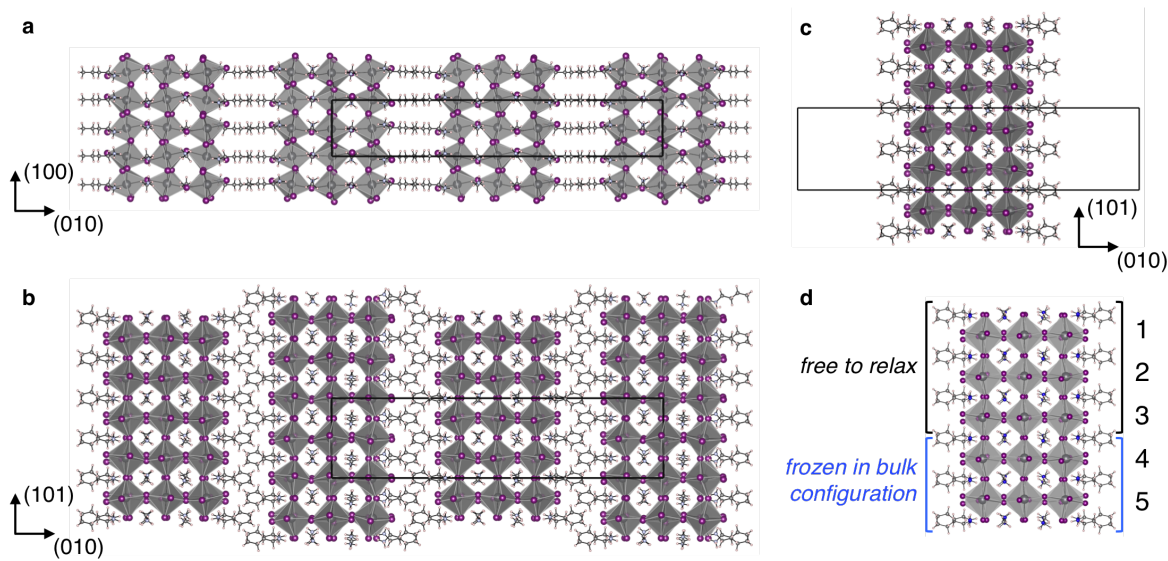

**Figure S4 | From bulk to (101) surface. a**, Modified structure of n=3 RPP where the original $Cmca$ orthorhombic cell is transformed to a tetragonal cell. The black line marks the unit cell of the modified bulk structures used in the calculation. **b**, Same structure reoriented along (101). The cell enlargement is limited by the tetragonal transformation. **c**, Only one inorganic layer is preserved, surrounded by 2 layers of the organic BA cations. **d**, A 5-octahedron thick slab is constructed. It is referenced as 'bulk-like' prior to relaxation. The bottom 2 layers of the slab are kept in bulk configuration, while the top 3 are allowed to relax freely, leading to the (101) surface. In all cases, Pb, I, N, C and H atoms are depicted in black, purple, blue, grey and pale pink, respectively.

**Table S2 |** Experimental ($a$, $c$) and modified ($a'=(a+c)/2$) norms of in-plane cell vectors for $(BA)_2(MA)_{n-1}Pb_nI_{3n+1}$ with n=1 to 4. $\eta$ gives the relative error caused by the distortion, *i.e.* a contraction of **a** and a dilatation of **c**.

| n | $a'$ (Å) | $a$ (Å) | $\eta$ (%) | $c$ (Å) | $\eta$ (%) |
|---|---|---|---|---|---|
| 1 | 8.784(4) | 8.876(4) | -1.04 | 8.692(5) | 1.06 |
| 2 | 8.903(0) | 8.947(0) | -0.49 | 8.858(9) | 0.50 |
| 3 | 8.902(6) | 8.927(5) | -0.28 | 8.877(7) | 0.28 |
| 4 | 8.904(5) | 8.927(4) | -0.26 | 8.881(6) | 0.26 |

## DFT and semi-empirical simulation of bulk $(BA)_2(MA)_{n-1}Pb_nI_{3n+1}$ RPP (n=1 to 4)

**Text II | Quantum and dielectric confinements and spatial overlap of wavefunctions**

As in strongly confined quantum wells, bulk $(BA)_2(MA)_{n-1}Pb_nI_{3n+1}$ RPPs photophysical properties are dominated by mobile electron-hole (e-h) pairs strongly bound by a Coulombic interaction (*i.e.*, Wannier-like excitons). Room-temperature spectroscopy gives access to the resulting exciton binding energy $E_b$.[5] It decreases from 380 meV and 270 meV for n=1 and 2, to 220 meV for n=3 and 4. Noteworthy, even for n=4, $E_b$ remains substantial,

3

preventing room-temperature efficient electron-hole-pair ionisation in absence of multiple interfaces or internal field in the device.

The strength of $E_b$ encompasses several physical phenomena that can be introduced gradually in a semi-empirical Bethe-Salpeter model.[6–8] In addition to quantum confinement, evidenced by flat bands along the stacking directions, bulk RPPs exhibit dielectric confinement stemming from the contrast of high-frequency dielectric constant $\epsilon_\infty(z)$ between the organic and perovskite layers.[8] The effect, strongest for n=1, decreases with increasing n. For n=4, $\epsilon_\infty(z)$ at the centre of the slab reaches almost the 3D MAPbI$_3$ value.[8] Taking into account quantum and dielectric confinements leads to a large overestimation of $E_b$.[8] However, the inclusion of wavefunctions (spatial expansion) in the evaluation of the exciton binding energy leads to excellent agreement between computed and experimental results. This evidences the dramatic effect of e-h overlap on the stability of the bound exciton states, which is used to further discuss exciton dissociation in the present work. Besides, ingredients of the model are obtained in the framework of DFT at a comparable level of theory than the one implemented here, thus supporting the validity of our approach.

## DFT investigation of (BA)$_2$(MA)$_{n-1}$Pb$_n$I$_{3n+1}$ surfaces

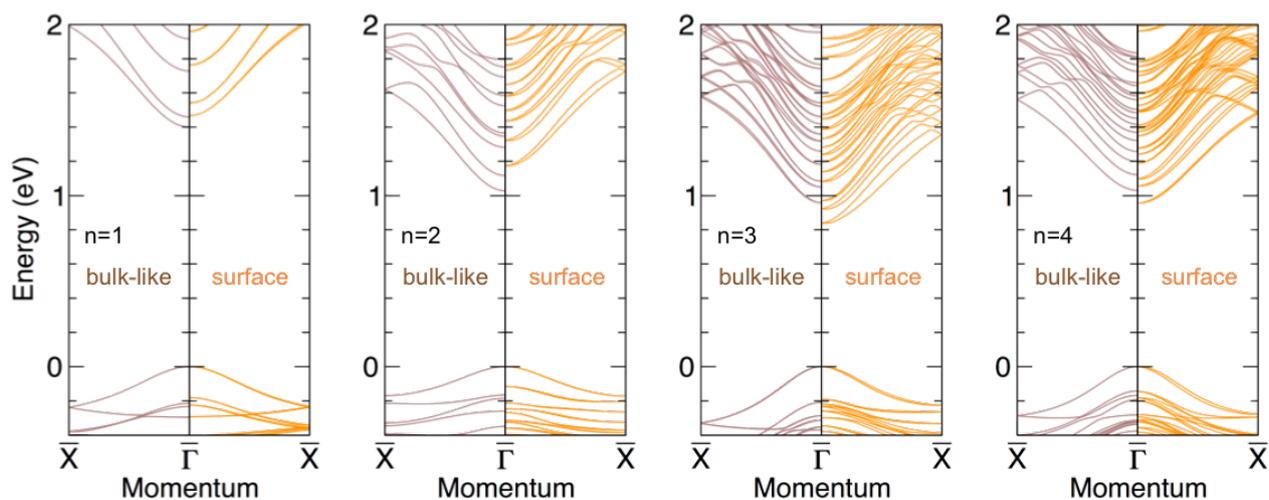

**Figure S5 | (101) surface band structures.** For each n value, the left panel shows the band structure of the bulk-like structure and the right panel the relaxed (101) surface band structure.

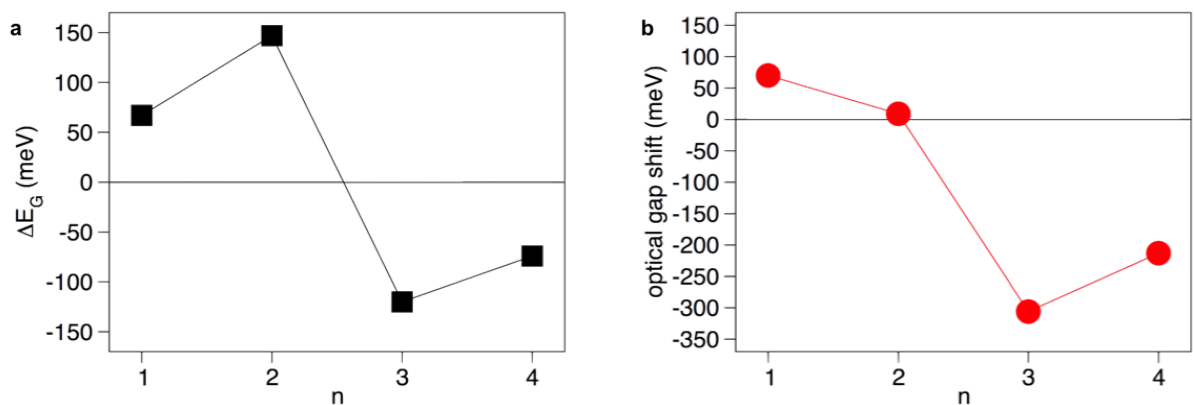

**Figure S6 | Gap shift. a** DFT variation of E$_G$ going from bulk-like to relaxed (101) surface. **b** Corresponding experimental shifts of the optical bandgap from exfoliated crystal to thin-film.[5]

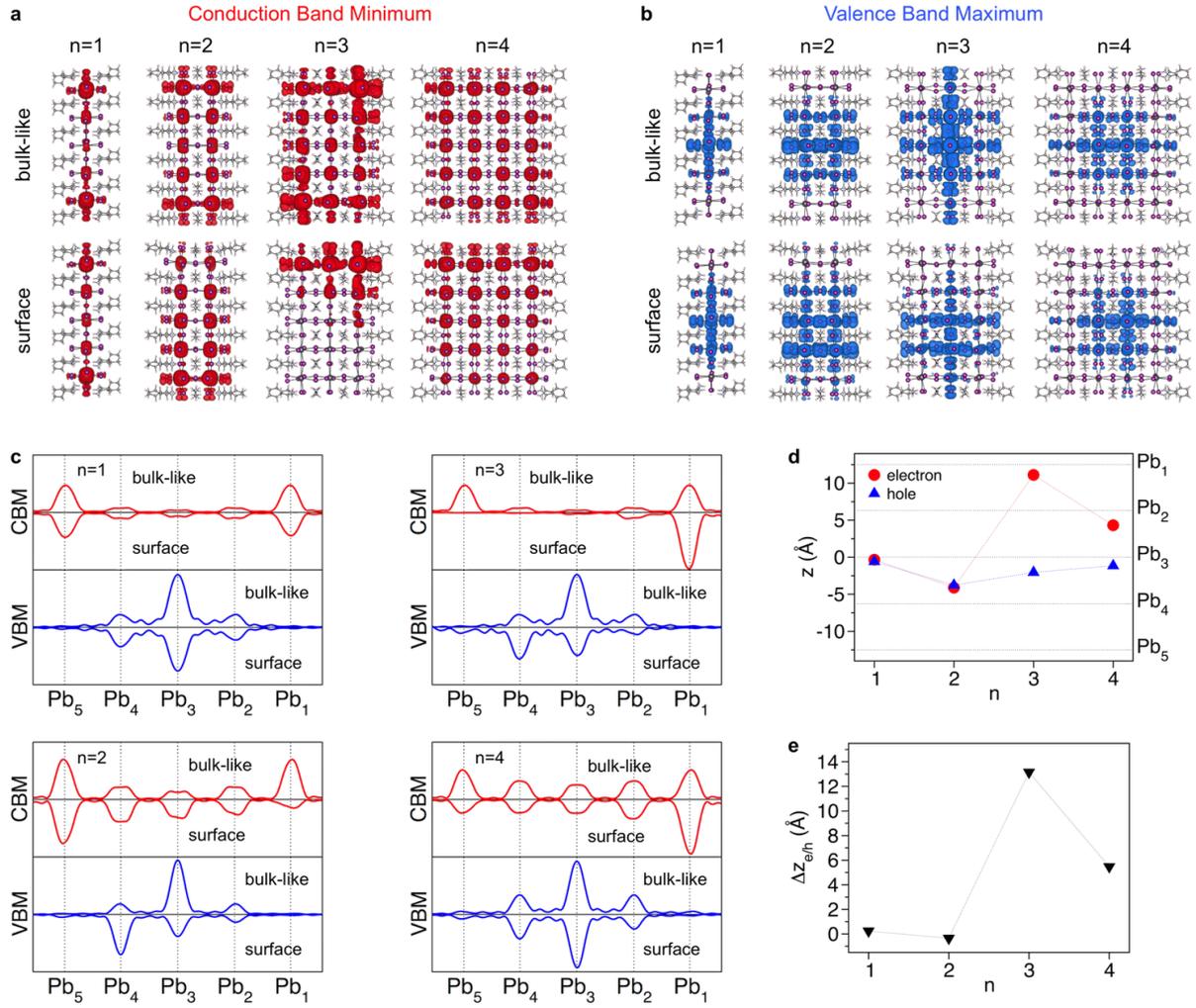

**Figure S7 | Electron and hole localization in (101) RPP surfaces. a**, LDOS at the conduction band minimum (CBM) in bulk-like and (101) surface slabs for n=1 to 4 RPPs. **b**, Same for the valence band maximum (VBM). **c**, LDOS profiles along the (101) direction for CBM and VBM of n=1 to 4 surfaces. The positions are indicated with respect to the five planes formed by lead atoms in the slab. $Pb_5$ points to the bottom plane that remains in bulk configuration, whereas $Pb_1$ is the top layer, *i.e.* the (101) surface. For n⩾3, the CBM gets localized at the surface with depletion in deeper regions. In all structures, the VBM tends to moves away from the surface. **d**, Position of the barycentres $z_e$ of electron (CBM) and $z_h$ of hole (VBM) states for the (101) slab. For all n, in bulk-like structures, the barycentres of CBM and VBM are always localized at the center of the slab. **e**, Electron-hole separation $\Delta z_{e/h}=z_e-z_h$ (in Å) for n=1 to 4 surfaces. For n⩽2, electron and holes overlap, whereas for n⩾3 a clear segregation occurs with a maximum separation for n=3 with $\Delta z_{e/h}>10$ Å.

**Table S3 | Kane energies (eV) computed for the n=3 RPP for bulk-like and (101) surface slabs.** As the conduction bands (CB) and valence bands (VB) of interest are doubly degenerated, we use the notation $CBM_{1,2}$, $VBM_{1,2}$, $CBM_{3,4}$ and $VBM_{3,4}$.

| | Bulk-like | | | | | | | |
|---|---|---|---|---|---|---|---|---|
| | $VBM_{1,2}$ | | | | $VBM_{3,4}$ | | | |
| | $x$ | $y$ | $z$ | norm | $x$ | $y$ | $z$ | norm |
| $CBM_{1,2}$ | 0.468 | 0.536 | 0.784 | 1.059 | 0.010 | 0.014 | 0.042 | 0.045 |
| $CBM_{3,4}$ | 0.737 | 0.870 | 0.732 | 1.355 | 0.166 | 0.212 | 0.072 | 0.279 |
| | Surface | | | | | | | |
| | $VBM_{1,2}$ | | | | $VBM_{3,4}$ | | | |
| | $x$ | $y$ | $z$ | norm | $x$ | $y$ | $z$ | norm |
| $CBM_{1,2}$ | 0.164 | 0.186 | 0.460 | 0.523 | 0.018 | 0.024 | 0.008 | 0.031 |
| $CBM_{3,4}$ | 0.086 | 0.116 | 0.134 | 0.197 | 0.010 | 0.006 | 0.010 | 0.015 |

# Design of layered hybrid perovskites

Table S4 | Chemical nomenclature and references for Fig. 4a.

| | Name | Formula | Ref. |
|---|---|---|---|
| BA | Butylammonium | CH$_3$(CH$_2$)$_3$NH$_3^+$ | 2 |
| PeA | Pentylammonium | CH$_3$(CH$_2$)$_4$NH$_3^+$ | 4 |
| HA | Hexylammonium | CH$_3$(CH$_2$)$_5$NH$_3^+$ | 4 |
| NoA | Nonylammonium | CH$_3$(CH$_2$)$_8$NH$_3^+$ | 9 |
| DoDA | Dodecylammonium | CH$_3$(CH$_2$)$_{11}$NH$_3^+$ | 10 |
| TeDA | Tetradecylammonium | CH$_3$(CH$_2$)$_{13}$NH$_3^+$ | 10 |
| HeDA | Hexadecylammonium | CH$_3$(CH$_2$)$_{15}$NH$_3^+$ | 10 |
| ODA | Octadecylammonium | CH$_3$(CH$_2$)$_{17}$NH$_3^+$ | 10 |
| HA$_2$ | 1,6-diammoniohexane | $^+$H$_3$N(CH$_2$)$_6$NH$_3^+$ | 11 |
| IEA | 2-iodoethylammonium | I-CH$_2$CH$_2$-NH$_3^+$ | 12 |
| NCEA | 2-cyanoethylammonium | N≡C-CH$_2$CH$_2$-NH$_3^+$ | 13 |
| cPeA | cyclopentylammonium | cyclo-C$_5$H$_9$-NH$_3^+$ | 14 |
| PhMA | phenylmethylammonium | C$_6$H$_5$-CH$_2$-NH$_3^+$ | 15 |
| 4MPhMA | 4-methylphenylmethylammonium | 4-CH$_3$-C$_6$H$_4$-CH$_2$-NH$_3^+$ | 16 |
| 1-PhEA | 1-phenylethylammonium | C$_6$H$_5$-CH$_2$CH$_2$-NH$_3^+$ | 17 |
| 4F-PhEA | 4-fluorophenylethylammonium | 4-F-C$_6$H$_4$-CH$_2$CH$_2$-NH$_3^+$ | 18 |
| 4Cl-PhA | 4-chlorophenylammonium | 4-Cl-C$_6$H$_4$-CH$_2$CH$_2$-NH$_3^+$ | 19 |
| 4Br-PhA | 4-bromophenylammonium | 4-Br-C$_6$H$_4$-CH$_2$CH$_2$-NH$_3^+$ | 20 |
| (ThEA)$_2$ | 5-ammoniumethylsulfanyl-2,2'-bithiophene | $^+$H$_3$N-CH$_2$CH$_2$-S-(thiophene)$_2$-S-CH$_2$CH$_2$-NH$_3^+$ | 21 |
| ThMA | 2-thienylmethylammonium | (thiophene)-CH$_2$-NH$_3^+$ | 22 |
| 2-hydroxyEA | 2-hydroxyethylammonium | HO-CH$_2$CH$_2$-NH$_3^+$ | 23 |
| HOOBA | 4-ammoniobutanoic | HOOC-(CH$_2$)$_3$-NH$_3^+$ | 24 |



\end{document}